\input{aipcheck}
\documentclass[
    ,final            
  ]
  {aipproc}

\layoutstyle{6x9}

\newcommand{\msun}{{M_{\odot}}}

\newcommand{\fmmt}{{\rm fm}^{-3}}

\newcommand{\ls}{{\stackrel{\textstyle <}{_\sim}}}

\begin{document}

\title{QCD in neutron stars and strange stars}


\author{F. Weber}{ address={Department of Physics, San Diego State
    University, 5500 Campanile Drive, San Diego, \\  CA 92182-1233, USA} }

\author{R. Negreiros}{ address={FIAS, Goethe University, Ruth Moufang
    Str.\ 1, 60438 Frankfurt, Germany}}


\begin{abstract}
  This paper provides an overview of the possible role of Quantum
  Chromo Dynamics (QDC) for neutron stars and strange stars. The
  fundamental degrees of freedom of QCD are quarks, which may exist as
  unconfined (color superconducting) particles in the cores of neutron
  stars. There is also the theoretical possibility that a
  significantly large number of up, down, and strange quarks may
  settle down in a new state of matter known as strange quark matter,
  which, by hypothesis, could be more stable than even the most stable
  atomic nucleus, $^{56}$Fe. In the latter case new classes of
  self-bound, color superconducting objects, ranging from strange
  quark nuggets to strange quark stars, should exist. The properties
  of such objects will be reviewed along with the possible existence
  of deconfined quarks in neutron stars. Implications for
  observational astrophysics are pointed out. 
\end{abstract}

\maketitle

\section{Introduction}

Astrophysicists distinguish between three types of compact
stars. These are white dwarfs, neutron stars, and black holes. Of the
three, neutron stars appear particularly interesting for QCD related
studies of ultra-dense matter, since the matter in the cores of such
objects is compressed to densities that are several times higher than
the densities of atomic nuclei
\cite{glen97:book,weber99:book,blaschke01:trento,baldo01:b,weber05:a,%
  haensel06:book,page06:review,klahn06:a_short,sedrakian07:a,klahn07:a}. Theoretical
studies indicate that at such densities hyperons may be generated and
new states of matter--such as boson condensates and/or quark
matter--may appear
\cite{glen97:book,weber99:book,blaschke01:trento,haensel06:book}. The
latter ought to be a color superconductor
\cite{rajagopal01:a,alford01:a,alford08:a}.  There is also the
intriguing theoretical possibility that strange quark matter could be
more stable than even the most stable atomic nucleus
\cite{witten84:a}, $^{56}$Fe, which would give rise to the existence
of new classes of compact objects, carrying baryon numbers ranging
from $\sim 10^2$ (quark nuggets) to $\sim 10^{57}$ (strange quark
stars)
\cite{weber05:a,farhi84:a,schaffner97:b,madsen98:a,schaffner06:a,alford06:a}.
This paper summarizes the role of QCD for neutron stars and strange
stars.  Particular emphasis is put on the role of
strangeness. Strangeness is carried by hyperons, quark matter, and
quark nuggets, and may leave its mark in the masses, radii, cooling
behavior, pycno-nuclear reactions, and the spin evolution of compact
stars.  \nobreak

\section{Limits on the central densities of neutron
  stars}\label{sec:variational}

Stringent limits on the central densities of neutron stars
\cite{hartle78:a,lattimer05:a} can be established through a
variational study of the poorly known nuclear equation of state
\cite{glen97:book,weber99:book,blaschke01:trento,haensel06:book,%
klahn06:a_short,klahn07:a}. Such
a study assumes that the equation of state of neutron star matter is
known up to an energy density $\epsilon_0$ and pressure $p_0$,
Einstein's theory of general relativity is the correct theory of
gravity, neutron star matter is microscopically stable (i.e.,
$\partial P/ \partial \epsilon >0$), and that causality if not
violated (i.e., $\partial P/ \partial \epsilon < 1$)
\cite{hartle78:a,glen92:limit}. Models for the nuclear equation of
state can then be generated from the following ansatz
\cite{glen92:limit},
\begin{eqnarray}
  \epsilon(u) = \frac{\alpha}{\gamma -1} (u^\gamma -u) +
  u\epsilon_0 + (1- u)(\alpha - p_0) \, , ~~ p(u) = \alpha(u^\gamma
  -1) + p_0 \, , \label{eq:var} 
\end{eqnarray}
where $p$ denotes pressure, $\epsilon$ stands for the energy density,
and $u \equiv \rho / \rho_0$ with $\rho$ the baryon number
density. The quantities $\alpha$ and $\gamma$ are parameters that
control the softness/stiffness of the nuclear equation of state. The
result of such a study \cite{weber10:iwara} show that the variational
upper limit on the masses of neutron stars is close to $2.9
\,\msun$. Such stars could have central densities up to four times
higher than the density of nuclear matter. Very recently the discovery
of a very massive neutron star, PSR~J1614-2230, was reported in
\cite{demorest10:a}.  This neutron star has a mass of $1.97\pm 0.04
\msun$ and rotates at 3.15 milliseconds, which, however, has only very
little impact on the star's structure. According to the study
presented above, the central density of this object could be anywhere
between $2 \ls \epsilon / \epsilon_0 \ls 10$ \cite{weber10:iwara}, the
high end of which evidently favoring the existence of exotic matter in
the core of this object (for a general discussion, see
\cite{miller10:a,oezel10:a}).

\section{Quark Matter in the Inner Cores of Neutron
  Stars}\label{sec:qm}

It has been suggested already many decades ago
\cite{ivanenko65:a,fritzsch73:a,baym76:a,keister76:a,%
  chap77:a,fech78:a,chap77:b} that the nucleons may melt under the
enormous pressure that exists in the cores of neutron stars, creating
a new state of matter known as quark matter. From simple geometrical
considerations it follows that for a characteristic nucleon radius,
$r_N$, of around one Fermi, nucleons may begin to touch each other in
nuclear matter at densities $(4\pi r^3_N/3)^{-1} \simeq 0.24~\fmmt =
1.5\, \rho_0$, which is less than twice the number density of nuclear
matter ($\rho_0=0.16~\fmmt$). This value increases to $\sim 11 \,
\rho_0$ for a nucleon radius of $r_N = 0.5$~fm. One may thus speculate
that the nucleons making up neutron star matter begin to dissolve at
densities somewhere between around $2-10\, \rho_0$, giving way to
quark matter made up of unconfined up and down quarks.  Depending on
rotational frequency and neutron star mass, densities greater than two
to three times $\rho_0$ may be easily reached in the cores of neutron
stars so that the neutrons and protons in the cores of neutron stars
may indeed be broken up into their quarks constituents
\cite{glen97:book,weber99:book,weber05:a,glen91:pt}. More than that,
since the mass of the strange quark is only $m_s \sim 150$~MeV,
high-energetic up and down quarks will readily transform to strange
quarks at about the same density at which up and down quark
deconfinement sets in. Thus, if quark matter exists in the cores of
neutron stars, it should be made of the three lightest quark flavors.
Possible astrophysical signals of quark deconfinement in the cores of
neutron stars were suggested in Refs.\
\cite{glen97:a,glen00:b,glen01:a,chubarian00:a,glen00:d}. They all
have their origin in the changes of the moment of inertia caused by
the gradual transformation of hadronic matter into quark matter, which
may lead to braking indices vastly different from the canonical value
of 3, to the spin-up of isolated rotating neutron stars for extended
(millions of years) periods of time, and to the pile-up of the
frequencies of (X-ray) neutron stars accreting matter from companion
stars.  If quark matter exists in neutron stars, it will consist of
the three lightest quark flavors only. Quarks carrying charm, top or
bottom flavors are much to massive to be generated in neutron stars
\cite{weber99:book,weber05:a}.

\section{Pycnonuclear Reactions}\label{sec:nuggets}

\begin{figure}[tb]
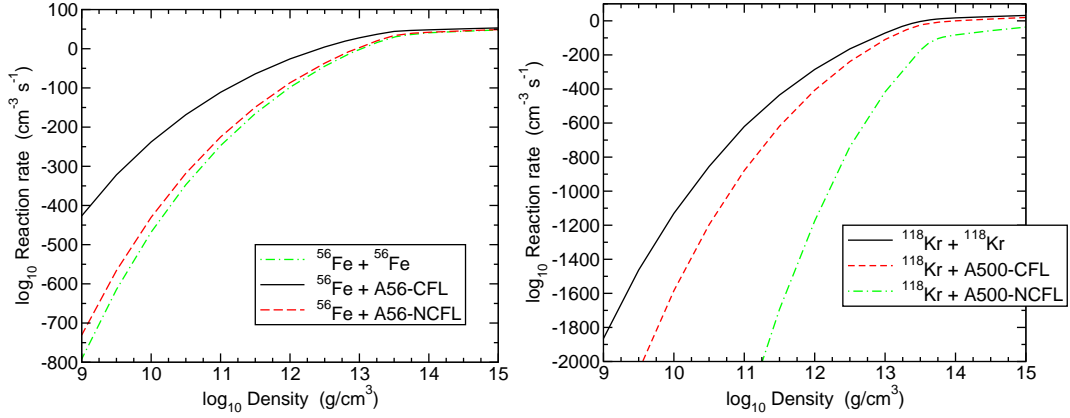

\begin{tabular}{cc}
  \includegraphics[height=0.25\textheight]{Fe56SQM.eps}
  \includegraphics[height=0.25\textheight]{Kr118LargeSQM.eps}
\end{tabular}
\caption{Nuclear reaction rates for $^{56}$Fe, $^{118}$Kr, and strange
  quark matter nuggets in the CFL/non-CFL (NCFL) phase with baryon
  numbers of 56 and 500 \cite{golf09:a}.}
\label{fig:pyc_rr}
\end{figure}

The lattice structures of the crusts of neutron stars makes these
regions suitable environments where pycnonuclear (fusion) reactions
among atomic lattice nuclei may occur
\cite{gasques05:a,yakovlev06:a}. Model calculations \cite{golf09:a}
indicate that the presence of strange quark matter nuggets could alter
these pycnonuclear reaction rates among the atomic lattice nuclei
tremendously, as shown in Fig.\ \ref{fig:pyc_rr}.
The differences in the reaction rates have their origin in the
different mass-to-charge ratios of strange quark matter nuggets and
atomic nuclei.  The calculations in Fig. \ref{fig:pyc_rr} are based on
the assumption that strange quark nuggets are made up of either
ordinary strange quark matter (NCFL) or color superconducting strange
quark matter whose condensation pattern is the color-flavor-locked
(CFL) phase. One crucial difference between non-CFL (NCFL) and CFL
quark matter is the equality of all quark Fermi momenta in CFL quark
matter which leads to charge neutrality in bulk without any need for
electrons \cite{rajagopal01:b}. This has most important consequences
for the charge-to-mass ratios of strangelets.  For ordinary (NCFL)
strangelets, the charge is approximately $Z \approx 0.1 m_{150}^2 A$
for $A\ll 10^3$, and $Z \approx 8 m_{150}^2 A^{1/3}$ for $A\gg 10^3$,
where $m_{150} \equiv m_s / 150\ {\rm MeV}$ and $m_s$ is the mass of
the strange quark. For small $A$, the charge is the volume quark
charge density multiplied by the strangelet volume with a result that
is proportional to $A$ itself. This relation holds until the system
grows larger than around 5 fm, or $A \approx 150$, at which point the
charge is mainly distributed near the strangelet surface, and $Z
\propto A^{1/3}$ \cite{madsen01:a}. In contrast to this, the
charge-to-mass ratio of CFL strangelets is described by $Z \approx
0.3\ m_{150} A^{2/3}$ \cite{madsen01:a} which leads to a
significantly increased pycnonuclear reaction rates in the crusts of
neutron stars, as shown in Fig.\ \ref{fig:pyc_rr}. Possible 
observational consequences concernt the thermal evolution of neutron stars
\cite{page05:a,stejner09:a} and maybe superbursts
\cite{page05:b,stejner06:a,cooper09:a}.

\section{Ultra-high electric fields and vortex expulsion}\label{sec:electric}

We now turn our attention to strange quark matter objects at the high
baryon number end ($A \sim 10^{57}$), also known as strange quark
stars. If existing, these objects would have masses and radii that are
similar to those of neutron stars, which makes it hard to distinguish
both types of stars from one another observationally. One of the major
differences between neutron stars and strange stars is that the latter
are self-bound objects so that very light but small (radii on the
order of just a few kilometers) strange quark stars could fill the
Universe.  Another striking feature of strange quark stars concerns
the existence of ultra-high electric fields a their surfaces
\cite{alcock86:a,alcock88:a,usov04:a,usov05:a}. This electric field is
a consequence of a high electron concentration near the stellar
surface, which is necessary to compensate the lower strange quark
population in this region, and to maintain electric charge
neutrality. As shown in
\cite{alcock86:a,alcock88:a,usov04:a,usov05:a}, these electrons are
screened out of the star and form an electric dipole layer with an
electric field on the order of $E \sim 10^{17-19}$ V/cm. Electric
fields of this magnitude can increase the stellar mass by up to
15\% \cite{negreiros09:a}. This is an important result since it allows
for the interpretation of massive pulsars as rotating strange stars.

The surface electric field can also give rise to differential rotation
of the star with respect to its surrounding electric surface field
\cite{negreiros10:diffrot}.  In this event electric currents are
generated at the surface of the strange star. The strength of these
currents is determined by the magnitude of the net electric charge and
by the amount of differential rotation.  The magnetic field of such a
configuration was found to be uniform inside the star, and of a dipole
type outside \cite{negreiros10:diffrot}. Moreover, depending on the
electric field and the relative frequency between the star and the
electron layer, the generated magnetic fields may be as high as
$10^{16}$~G. Such strong fields can be achieved for very high static
electric fields on the order of $\sim 10^{20}-10^{21}$~V/cm and
effective frequencies of $\sim 700 - 1000$~Hz. For small effective
rotational frequencies of $\sim 10$~Hz and more moderate static
electric fields of $\sim 10^{16}-10^{18}$~V/cm one obtains magnetic
fields on the order of $10^{9}-10^{11}$~G. This is a very intriguing
result because such magnetic fields and rotational frequencies are in
good agreement with the observed magnetic fields and frequencies of
three Central Compact Objects (CCOs) \cite{negreiros10:diffrot}.  CCOs
form a group of recently discovered compact stars that are
characterized by a faint steady flux predominately in the X-ray range
and the absence of optical and radio counterparts
\cite{becker09:a,halpern10:a}.  CCOs have relatively long rotational
periods and, for the three cases for which data exists, possess small
magnetic fields of $\sim 10^{11}$~G \cite{halpern10:a}. These objects
could thus be comfortably interpreted as rotating strange stars whose
electron atmospheres rotate at frequencies that a slightly different
from the ones of the stellar cores. The scenario described just above
is only for strange stars made of color-flavor locked (CFL)
superconducting quark matter but not for two-flavor color
superconducting (2SC) quark matter, which has very different
properties \cite{alford08:a}.

Strange stars made of CFL quark matter ought to be threaded with
rotational vortex lines within which the star's interior magnetic
field is confined. If so, the vortices (and thus magnetic flux) would
be expelled from the star during stellar spin-down, leading to
magnetic reconnection at the surface of the star and the prolific
production of thermal energy \cite{ouyed04:a}.  In
\cite{niebergal10:a} it was shown that this energy release can re-heat
quark stars to exceptionally high temperatures, such as observed for
Soft Gamma Repeaters (SGRs), Anomalous X-Ray pulsars (AXPs), and X-ray
dim isolated neutron stars (XDINs). Moreover, numerical investigations
\cite{niebergal10:a} of the temperature evolution, spin-down rate, and
magnetic field behavior of such superconducting quark stars suggest
that SGRs, AXPs, and XDINs may be linked ancestrally.  Finally, the
density at which quarks deconfine follows from this study to be of the
order of five times that of nuclear saturation density, which is well
within reach of typical neutron star densities
\cite{glen97:book,weber99:book,blaschke01:trento,baldo01:b,%
haensel06:book,page06:review,klahn06:a_short,sedrakian07:a,klahn07:a}.

\section{Conclusions}\label{sec:conclusions}

The purpose of this short paper is to provide an overview of the
multifaceted role of QCD for compact stars.  We began with an
investigation of the maximum densities of neutron stars.  The results
indicate that even very massive ($\sim 2\, \msun$) neutron stars can
have tremendous central densities (up to 10 time nuclear), which
leaves plenty of leeway for the possible existence of hyperons, boson
condensates, and/or deconfined up, down and strange quarks in the
cores of such objects. Depending on the details of a possible
deconfinement phase transition in the cores of neutron stars,
anomalies in the rotational evolution (backbending) may be triggered
by the transition which could be observed by radio and X-ray
telescopes. If strange quark matter were absolutely stable, nuggets
made of strange quark matter could exit in the crusts of neutron
stars. If in the CFL phase, he presence of such nuggets may
tremendously increase the pycnonuclear reaction rates in the crusts of
neutron stars, which may serve as an observational window on the
actual existence of strange quark matter. Stars made of strange quark
matter could possess huge electric fields, increasing the stellar
mass by up to 15\%. This is an important result since it facilitates
the interpretation of massive pulsars as rotating strange quark
stars. Finally it was pointed out that CFL superconducting strange stars
may distinguish themselves from ordinary neutron stars by
differentially rotating electron surface layers, which could explain
the magnetic fields observed for several compact central objects, and
the possibility of vortex expulsion of magnetic flux lines from
the star, which leads to a significant reheating of such stars.
The computed temperatures are in excellent agreement with
those observed for those of magnetars.

\begin{theacknowledgments}
  This work is supported by the National Science Foundation under
  Grant PHY-0854699.
\end{theacknowledgments}


\begin{thebibliography}{10}

\bibitem{glen97:book}
N. K. Glendenning, {\it Compact Stars, Nuclear Physics, Particle Physics, and
  General Relativity}, 2nd ed.\ (Springer-Verlag, New York, 2000).

\bibitem{weber99:book}
F. Weber, {\it Pulsars as Astrophysical Laboratories for Nuclear and Particle
  Physics}, High Energy Physics, Cosmology and Gravitation Series (IOP
  Publishing, Bristol, Great Britain, 1999).

\bibitem{blaschke01:trento}
{\it Physics of Neutron Star Interiors}, ed.\ by D.\ Blaschke, N.\ K.\
  Glendenning, and A.\ Sedrakian, Lecture Notes in Physics {\bf 578}
  (Spring-Verlag, Berlin, 2001).

\bibitem{baldo01:b}
M. Baldo and F. Burgio, {\it Microscopic Theory of the Nuclear Equation of
  State and Neutron Star Structure}, Lecture Notes in Physics {\bf 578},
  (Springer-Verlag, Berlin, 2001), p.\ 1.

\bibitem{weber05:a}
F. Weber, Prog.\ Part.\ Nucl.\ Phys.\ {\bf 54} (2005) 193.

\bibitem{haensel06:book}
P. Haensel, A. Y. Potekhin, and D. G. Yakovlev, {\it Neutron Stars 1},
  Astrophysics and Space Science Library, (Springer-Verlag, New York, 2006).

\bibitem{page06:review}
D. Page and S. Reddy, Ann.\ Rev.\ Nucl.\ Part.\ Sci.\ {\bf 56} (2006) 327.

\bibitem{klahn06:a_short}
T. Kl{\"{a}}hn {\it et al.}, Phys.\ Rev.\ C {\bf 74} (2006) 035802.

\bibitem{sedrakian07:a}
A. Sedrakian, Prog.\ Part.\ Nucl.\ Phys.\ {\bf 58} (2007) 168.

\bibitem{klahn07:a}
T. Kl{\"{a}}hn {\it et al.}, Phys.\ Lett.\ B {\bf 654} (2007) 170.

\bibitem{rajagopal01:a}
K. Rajagopal and F. Wilczek, {\it The Condensed Matter Physics of QCD}, At the
  Frontier of Particle Physics/Handbook of QCD, ed.\ M.\ Shifman, (World
  Scientific, 2001).

\bibitem{alford01:a}
M. Alford, Ann.\ Rev.\ Nucl.\ Part.\ Sci.\ {\bf 51} (2001) 131.

\bibitem{alford08:a}
M. G. Alford, A. Schmitt, K. Rajagopal, and T. Sch{\"{a}}fer, Rev.\ Mod.\
  Phys.\ {\bf 80} (2008) 1455.

\bibitem{witten84:a}
E. Witten, Phys.\ Rev.\ D {\bf 30} (1984) 272.

\bibitem{farhi84:a}
E. Farhi and R. L. Jaffe, Phys.\ Rev.\ D {\bf 30} (1984) 2379.

\bibitem{schaffner97:b}
J. Schaffner-Bielich, C. Greiner, A. Diener, and H. St{\"{o}}cker, Phys.\ Rev.\
  C {\bf 55} (1997) 3038.

\bibitem{madsen98:a}
J. Madsen, Phys.\ Rev.\ Lett.\ {\bf 81} (1998) 3311.

\bibitem{schaffner06:a}
I. Sagert, M. Wietoska, J. Schaffner-Bielich, J. Phys. G {\bf 32} (2006) S241.

\bibitem{alford06:a}
M. Alford, K. Rajagopal, S. Reddy, and A. W. Steiner, Phys.\ Rev.\ D {\bf 73}
  (2006) 114016.

\bibitem{hartle78:a}
J. B. Hartle, Phys.\ Rep.\ {\bf 46} (1978) 201.

\bibitem{lattimer05:a}
J. Lattimer and M. Prakash, Phys.\ Rev.\ Lett.\ {\bf 94} (2005) 111101.

\bibitem{glen92:limit}
N. K. Glendenning, Phys.\ Rev.\ D {\bf 46} (1992) 4161.

\bibitem{weber10:iwara}
F.Weber, O. Hamil, K. Mimura, and R. Negreiros, IJMP 19 (2010) 1427.

\bibitem{demorest10:a}
P. B. Demorest, T. Pennucci, S. M. Ransom, M. S. E. Roberts and J. W. T.
  Hessels, Nature {\bf 467} (2010) 1081.

\bibitem{miller10:a}
M. Coleman Miller, Nature {\bf 467} (2010) 1057.

\bibitem{oezel10:a}
F. {\"{O}}zel, D. Psaltis, S. Ransom, P. Demorest, and M. Alford, Astrophys.\
  J.\ {\bf 724} (2010) L199.

\bibitem{ivanenko65:a}
D. D. Ivanenko and D. F. Kurdgelaidze, Astrophys.\ {\bf 1} (1965) 251.

\bibitem{fritzsch73:a}
H. Fritzsch, M. Gell--Mann, and H. Leutwyler, Phys.\ Lett.\ {\bf 47B} (1973)
  365.

\bibitem{baym76:a}
G. Baym and S. Chin, Phys.\ Lett.\ {\bf 62B} (1976) 241.

\bibitem{keister76:a}
B. D. Keister and L. S. Kisslinger, Phys.\ Lett.\ {\bf 64B} (1976) 117.

\bibitem{chap77:a}
G. Chapline and M. Nauenberg, Phys.\ Rev.\ D {\bf 16} (1977) 450.

\bibitem{fech78:a}
W. B. Fechner and P. C. Joss, Nature {\bf 274} (1978) 347.

\bibitem{chap77:b}
G. Chapline and M. Nauenberg, Ann.\ New York Academy of Sci.\ {\bf 302} (1977)
  191.

\bibitem{glen91:pt}
N. K. Glendenning, Phys.\ Rev.\ D {\bf 46} (1992) 1274.

\bibitem{glen97:a}
N. K. Glendenning, S. Pei, and F. Weber, Phys.\ Rev.\ Lett.\ {\bf 79} (1997)
  1603.

\bibitem{glen00:b}
N. K. Glendenning and F. Weber, {\it Signal of Quark Deconfinement in
  Millisecond Pulsars and Reconfinement in Accreting X-ray Neutron Stars},
  Lecture Notes in Physics {\bf 578}, (Springer-Verlag, Berlin, 2001), p.\ 305.

\bibitem{glen01:a}
N. K. Glendenning and F. Weber, Astrophys.\ J.\ {\bf 559} (2001) L119.

\bibitem{chubarian00:a}
E. Chubarian, H. Grigorian, G. Poghosyan, and D. Blaschke, Astron.\ {\&}
  Astrophys.\ {\bf 357} (2000) 968.

\bibitem{glen00:d}
N. K. Glendenning and F. Weber, {\it Spin Clustering as Possible Evidence of
  Quark Matter in Accreting X-ray Neutron Stars}, AIP conf.\ proc.\ {\bf 610}
  (2002) p.\ 470.

\bibitem{gasques05:a}
L. R. Gasques, A. V. Afanasjev, E. F. Aguilera, M. Beard, L. C. Chamon, P.
  Ring, M. Wiescher, and D. G. Yakovlev, Phys.\ Rev.\ C {\bf 72} (2005) 025806.

\bibitem{yakovlev06:a}
D. G. Yakovlev, L. R. Gasques, M. Beard, M. Wiescher, and A. V. Afanasjev,
  Phys.\ Rev.\ C {\bf 74} (2006) 035803.

\bibitem{golf09:a}
B. Golf, J. Hellmers, and F. Weber, Phys.\ Rev.\ C {\bf 80} (2009) 015804.

\bibitem{rajagopal01:b}
K. Rajagopal and F. Wilczek, Phys.\ Rev.\ Lett.\ {\bf 86} (2001) 3492.

\bibitem{madsen01:a}
J. Madsen, Phys.\ Rev.\ Lett.\ {\bf 87} (2001) 172003.

\bibitem{page05:a}
D. Page, U. Geppert, and F. Weber, Nucl. Phys. {\bf A 777} (2006) 492.

\bibitem{stejner09:a}
M. Stejner, F. Weber, and J. Madsen, Astrophys.\ J. {\bf 694} (2009) 1019.

\bibitem{page05:b}
D. Page and A. Cumming, Astrophys.\ J.\ {\bf 635} (2005) L157.

\bibitem{stejner06:a}
M. Stejner and J. Madsen, Astron.\ Astrophys.\ {\bf 458} (2006) 523.

\bibitem{cooper09:a}
R. L. Cooper, A. W. Steiner, and E. F. Brown, Astrophys.\ J.\ {\bf 702} (2009)
  660.

\bibitem{alcock86:a}
C. Alcock, E. Farhi, and A. V. Olinto, Astrophys.\ J.\ {\bf 310} (1986) 261.

\bibitem{alcock88:a}
C. Alcock and A. V. Olinto, Ann.\ Rev.\ Nucl.\ Part.\ Sci.\ {\bf 38} (1988)
  161.

\bibitem{usov04:a}
V. V. Usov, Phys.\ Rev.\ D {\bf 70} (2004) 067301.

\bibitem{usov05:a}
V. V. Usov, T. Harko, and K. S. Cheng, Astrophys.\ J.\ 620 (2005) 915.

\bibitem{negreiros09:a}
R. Negreiros, F. Weber, M. Malheiro, and V. Usov, Phys.\ Rev.\ D {\bf 80}
  (2009) 083006.

\bibitem{negreiros10:diffrot}
R. P. Negreiros, I. N. Mishustin, S. Schramm, and F. Weber, PRD {\bf 82} (2010)
  103010.

\bibitem{becker09:a}
W. Becker, and J. Truemper, ASSL {\bf 357} (2009) 91.

\bibitem{halpern10:a}
J. P. Halpern and E. V. Gotthelf, Astrophys. \ J. {\bf 709} (2010) 436.

\bibitem{ouyed04:a}
R. Ouyed, {\O}. Elgar{\o}y, H. Dahle, and P. Ker{\"{a}}nen, Astron. {\&}
  Astrophys.\ {\bf 420} (2004) 1025.

\bibitem{niebergal10:a}
B. Niebergal, R. Ouyed, R. Negreiros, and F. Weber, PRD {\bf 81} (2010) 043005.

\end{thebibliography}

\bibliographystyle{aipproc}   

\end{document}